# Observation of Half Quantum Vortices in an Exciton-Polariton Condensate

K. G. Lagoudakis[1], T. Ostatnický[2], A.V. Kavokin[2,3], Y. G. Rubo[4], R. André[5], B. Deveaud-Plédran[1]

[1]IPEQ, École Polytechnique Fédérale de Lausanne(EPFL),1015 Lausanne, Switzerland.

[2]School of Physics and Astronomy, University of Southampton, Highfield, Southampton SO17 1BJ, United Kingdom.

[3]Marie-Curie Chair of Excellence, University of Rome II, 1, via della Ricerca Scientifica, Rome, 00133, Italy.

[4]Centro de Investigación en Energía, Universidad Nacional Autónoma de México, Temixco, Morelos, 62580, Mexico.

[5]Institut Néel, CNRS, Grenoble, France.

**Singly quantized vortices have been already observed in many systems including the superfluid helium, Bose Einstein condensates of dilute atomic gases, and condensates of exciton polaritons in the solid state. Two dimensional superfluids carrying spin are expected to demonstrate a different type of elementary excitations referred to as half quantum vortices characterized by a $\pi$ rotation of the phase and a $\pi$ rotation of the polarization vector when circumventing the vortex core. We detect half quantum vortices in an exciton-polariton condensate by means of polarization resolved interferometry, real space spectroscopy and phase imaging. Half quantum vortices coexist with single quantum vortices in our sample.**

Quantized vorticity is a property of quantum fluids that has been widely studied in various types of superfluids either with or without spin (*1- 4*). Superfluids with a two component (spinor) order parameter are characterized by a different type of vortices than those found in conventional scalar superfluids (*5- 8*). This new type of vortices is the so called half quantum vortices (HQV). They have lower energy with respect to full vortices and constitute the elementary excitations of spinor superfluids. When circumventing their core, the phase and the polarization vector experience a $\pm\pi$ rotation. In this sense, HQV can be understood as a half phase vortex combined with a half polarization vortex (*5*). In $^4$He superfluids the HQV cannot be formed as the spin degree of freedom is absent. However, in $^3$He (*5*, *6*) or in triplet superconductors (*7*, *8*) the order parameter has two or more components, so that the formation of HQV is possible. So far, experiments have not presented an unambiguous evidence for HQV in $^3$He (*9*), while more reliable indications of existence of HQV have been reported in cuprate superconductors (*10*, *11*). Recently, HQV have been proposed as a smoking gun for the superfluid of exciton-polaritons in semiconductor microcavities (*12*). We report on a direct observation of HQVs in a high temperature quantum fluid: microcavity exciton-polaritons. Exciton-polaritons are composite bosons carrying spin. They can occur in semiconductor microcavities in the strong coupling regime and are partly excitons and partly photons. The energy, wavevector, polarization and statistics of cavity polaritons are directly transferred to light emitted by the microcavity due to photon tunneling through the mirrors of the cavity (*13*). Due to their photonic component, the cavity polaritons have an extremely low effective mass of the order of $10^{-4}$ the free electron mass which allows for their Bose Einstein condensation (BEC) at temperatures achievable by cryogenic means. BEC of exciton polaritons has been demonstrated in various types of microcavities composed either of II-VI (*14*) or III-V compounds (*15*). The most prominent effects found in these systems are the bosonic stimulation, the appearance of long range spatial coherence and the build up of the vector polarization (*13*). Several indications of the polariton superfluidity have been reported including the Bogoliubov like dispersion (*16*), the appearance of phase vortices (*4*) and diffusionless motion of coherent polariton fluids in the presence of obstacles (*17*). In this work, HQVs are reported in a II-VI microcavity where BEC of the exciton-polaritons and formation of the integer phase vortices have been reported recently (*4, 14*). Coexistence of HQV and integer vortices in the same sample is possible because of the spatially inhomogeneous polarization splitting in microcavities, which is responsible for the suppression of HQV in certain areas of the sample. Thus, in different parts of the sample the polariton fluid has a different topology.

In order to fully characterize a vortex in a polariton condensate, one needs two winding numbers, (*k,m*), one for the polarization angle and one for the phase. One can express the order parameter of the condensate in the linear *xy*-basis as

$$\psi_{lin}(\mathbf{r}) = \sqrt{n} e^{i\theta(\mathbf{r})} \begin{pmatrix} \cos\eta(\mathbf{r}) \\ \sin\eta(\mathbf{r}) \end{pmatrix} \tag{1}$$

where $\theta(\mathbf{r})$ is the phase of the coherent polariton fluid and $\eta(\mathbf{r})$ is the polar angle that characterizes the orientation of the electric field of polaritons, i.e. the polarization angle. Vortices are described in this notation by rotation of the phase and the polarization as $\eta(\mathbf{r}) \to \eta(\mathbf{r}) + 2\pi k$ and $\theta(\mathbf{r}) \to \theta(\mathbf{r}) + 2\pi m$ where *k,m* can take integer or half integer values with $(k+m) \in \mathbb{Z}$. Four types of half vortices are described by winding numbers $(k,m) = \left(\pm\frac{1}{2}, \pm\frac{1}{2}\right)$. In order to reveal the specific phenomenology of HQVs with respect to the integer vortices it is convenient to analyze the circularly polarized components of the order parameter, which can be expressed as

$$\psi_{lin}(\mathbf{r}) = \frac{\sqrt{n}}{2}\left[ e^{i(\theta(\mathbf{r})-\eta(\mathbf{r}))} \begin{pmatrix} 1 \\ i \end{pmatrix} + e^{i(\theta(\mathbf{r})+\eta(\mathbf{r}))} \begin{pmatrix} 1 \\ -i \end{pmatrix} \right] \tag{2}$$

One can see that for $\eta(\mathbf{r}) \to \eta(\mathbf{r}) + \pi$ and $\theta(\mathbf{r}) \to \theta(\mathbf{r}) + \pi$ a zero rotation takes place for one circular polarization and a full $2\pi$ rotation is achieved for the other circular polarization. This means that if one were to detect a half vortex, it would be easiest when looking in $\sigma^+$ and $\sigma^-$ polarizations simultaneously. Then HQV would be observed as a full vortex in one polarization and no vortex in the other one. A signature for the phase vortex is a forklike dislocation in the interference pattern (*4, 18*). In the case of full phase vortices the forklike dislocations are expected to be seen in the same place in both circular polarizations, while in the case of HQV the fork appears only in one of the circular polarizations. In the circular basis one can write the order parameter of HQV in cylindrical coordinates as

$$\psi_{k,m}(\rho,\phi) = \sqrt{\frac{n}{2}} e^{im\phi} \begin{Bmatrix} [f(\rho) + \text{sgn}(km) \ g(\rho)] \cdot e^{-ik\phi} \\ [f(\rho) - \text{sgn}(km) \ g(\rho)] \cdot e^{ik\phi} \end{Bmatrix} \quad (3)$$

with $\rho = \frac{r}{a}$ being the relative distance from the vortex core in vortex radii and $\phi$ being the angular coordinate. The form of the two radial density functions *f* and *g* is known (*12*) and will give zero density for one circular polarization (*f - g*) and a finite density for the other polarization (*f + g*), as it is expected for the simplistic image of a full vortex in one circular polarization and no singularities for the other circular polarization.

An important feature of polariton condensates is the presence of polarization splitting induced by the structural anisotropy and stationary disorder. This splitting pins the polarization vector of the condensate to a given crystal axis. It is theoretically predicted that HQVs still exist in this case but the spatial distribution of the polariton vector field is modified. Similarly to the vortices in multi-component quantum Hall systems (*19*), the polariton half vortices acquire "strings" (or solitons, (*6,* Fig. 16.1)), whereby the polarization angle rotates by $\pi$ (*20*). The width of the string is given by $\hbar/\sqrt{2m^*\varepsilon}$, where $m^*$ is the effective mass of polaritons and $\varepsilon$ is the energy of the polarization splitting. HQVs remain the lowest energy topological defects if this width is greater or comparable to the excitation spot radius. However when this length becomes comparable to the vortex core size ($a \approx \hbar/\sqrt{2m^*\mu}$, where $\mu$ is the chemical potential), the excitation of HQVs would require too much energy and the integer phase vortices

$(0,\pm 1)$ become elementary topological excitations. For a realistic vortex core size of the order of ~2μm and the polariton mass $m^* \approx 10^{-4} m_e$, pairs of HQVs will be replaced by integer phase vortices for polarization splittings $\varepsilon \geq 100 \mu eV$. (Two close pairs of HQVs are shown in (*21*), Figs. S2, S3).

The situation in real microcavity samples is additionally complicated by the fact that the polarization splitting $\varepsilon$ fluctuates as a function of the coordinate in the plane of the cavity. This is why the HQV and integer phase vortices may coexist within the same condensate. The underlying mechanisms for the polarization splitting are thought to be the different penetration depths in the distributed Bragg reflectors (microcavity mirrors) for TE and TM polarizations (*22*) and the intrinsic anisotropy of the microcavity (*23*, *24*). The anisotropy is expected to be the product of a number of parameters, including the alloy concentrations, the wedge, QW width fluctuations and the built-in strain. Splittings vary from zero to several tenths of μeVs. All HQVs that we observed in this sample were at regions where the splitting was less than our experimental resolution ($\approx 20 \mu eV$).

The sample we studied is the same CdTe/CdMgTe microcavity that was used in our previous experiments cooled down to ~10 K by a liquid helium flow cryostat (*4*). We used continuous wave monomode non-resonant optical excitation. Detection was performed by means of the modified Michelson interferometer with active stabilization (*4*) completed by a lambda quarter and a Wollaston prism, to allow for polarization resolved interferometry in $\sigma^+$ and $\sigma^-$ polarizations simultaneously, which facilitated the identification of half vortices (*21*). All spectral studies were performed using a monochromator with $\approx 20 \mu eV$ resolution. The output of the interferometer could be sent to the entrance slits of the spectrometer through a polarizer, allowing for spectrally and polarization resolved interferometry images to be acquired. The HQV were observed only at the excitation powers exceeding the condensation threshold. Once a good candidate was found, then we performed a number of preliminary "test experiments" to verify unambiguously the persistence of the vortex for all possible detection configurations. The two most reliable tests were to change the overlap conditions at the output of the interferometer by shifting the mirror arm image with respect to one reflected from the retroreflector and to rotate by π the orientation of the fringes, making sure that for all orientations the singularity of the vortex is always clearly observable (*4*). We took care to verify the mutual coherence of the two cross circular polarization components by means of polarization mixing interferometry in order to eliminate the possibility of having two independent condensates in the two polarizations. In all cases we observed excellent mutual coherence properties with good contrast in the interference fringes between the two circular polarization components (*21*). The appearance of half vortices was quite rare, that is one out of six regions with no polarization splitting was exhibiting a HQV.

Once the HQV was identified, the interferometric image was being sent on the entrance slits of the spectrometer. Then we performed an optical tomography experiment (*25*) for $\sigma^+$ and $\sigma^-$ polarized images which provided us with the full set of polarization resolved interferograms in real space for all frequencies within the observable spectral window. Fig. 1(A,B) shows the reconstructed interferogram coming from the frequency of the polariton condensate for the polarizations $\sigma^+$ and $\sigma^-$, on which we have added a red circle at the center of the vortex core to help the reader locating the singularity. The singularity (forklike dislocation) is clearly visible for the $\sigma^+$ polarization whereas on the same position in real space for the $\sigma^-$ polarization we observe straight fringes. The interference patterns gave us access to the phase of the coherent polariton fluid. To extract the phase we assumed that the reference field coming from a region of the condensate without a vortex has a flat phase profile. Fig. 1(C,D) shows the phase of the polariton fluid in real space calculated from the interferograms. The phase has distinguishable characteristics only where there is enough signal intensity, whereas at the regions with no signal we get a noisy phase with no distinguishable features. The position of the HQV in the phase map is highlighted by circles. In $\sigma^+$ polarization, the phase rotates by 2π as one goes around the core. This behavior of the phase is clearly seen within an area of a few microns size. In the same region for the $\sigma^-$ polarization there are no observable singularities and the phase is homogeneous. Fig. 1(E,F) shows the phase as a function of the azimuthal angle as one goes around the core along the circles of different radii (shown by color). For the radius of 1μm, the phase changes monotonously in $\sigma^+$ polarization decreasing by 2π as one makes a full round. Contrary to this, for the same radius in $\sigma^-$ polarization we observe a quasi-flat phase profile indicating the lack of any singularity. For larger radii, the phase dependence on the azimuthal angle becomes strongly non-linear, while the total phase shift as one goes around the core remains -2π for one and 0 for the other polarization. Distortion of the phase profile at the large radii may be indicative for the existence of nearby regions with substantial vorticity, but can also be indicative for formation of a string.

We note that the specific HQV shown in Fig. 1 is characterized by the winding numbers $(k,m) = \left(+\frac{1}{2}, -\frac{1}{2}\right)$, while we have observed also the three remaining types of HQVs in different locations on the sample (21). On the basis of measurements we have done, we believe that four possible types of HQV are realized with approximately the same probability in our sample.

Using the same tomographic technique of spectrally resolved real space imaging as before, we then probed only the density of polaritons in the condensate state (Fig. 2A,B). Locating the vortex in real space and looking at the density close to its core, we observed that a local minimum for one polarization coincides with a maximum for the other one, as Fig. 2(C,D) shows. The widths of these minima/maxima coincide with the theoretical vortex core size $a$. This behavior is another signature of HQVs as one can see from Eq. (3). The theory (12) predicted that at the center of the HQV the condensate should be fully circularly polarized, and this is exactly what we observe in Fig. 2(C,D).

The HQVs we observed here are pinned by the disorder to specific locations on the sample. This is confirmed by the behavior of the interferometric images as a function of the pumping power. Increasing the excitation intensity, we modify the effective disorder potential acting upon the polariton condensate by changing the polariton-polariton repulsion strength. When pumping strongly enough, we screen the disorder potential so that HQV get unpinned and disappear from the interferometry image of a specific spot on the sample. This is what we observed at the excitation power exceeding the threshold pumping by a factor of 4.5. Above this power the forklike dislocation in $\sigma^+$ polarization disappears (*21*).

The stationary disorder fixes the winding numbers of the pinned vortices, so that repeating the experiment we find HQVs with the same winding numbers in the same locations. This is also true for the integer vortices. Handedness of each pinned vortex is dependent on the direction of polariton fluxes propagating in the disorder landscape during formation of the condensate, as the modeling based on the Gross-Pitaevskii equation showed (*4,26*).

This experimental work provides direct evidence of half quantum vortices in a spinor condensate, by means of polarization resolved interferometry, phase imaging and spectrally resolved real space density imaging (*27*).

27. We thank Le Si Dang for crucial remarks and detailed proofreading of this manuscript. We also thank T.C.H. Liew, M. Toledo Solano, M. Wouters, B. Pietka, Y. Leger, M. Richard, A. Baas, G. Nardin, D. N. Krizhanovskii, D. Sanvitto, P. G. Lagoudakis and V. Savona, for enlightening discussions. K. G. Lagoudakis and B. Deveaud-Plédran thank QP-NCCR for financial support through SNSF. A. Kavokin and Y.G. Rubo thank EPSRC and DGAPA-UNAM for support.

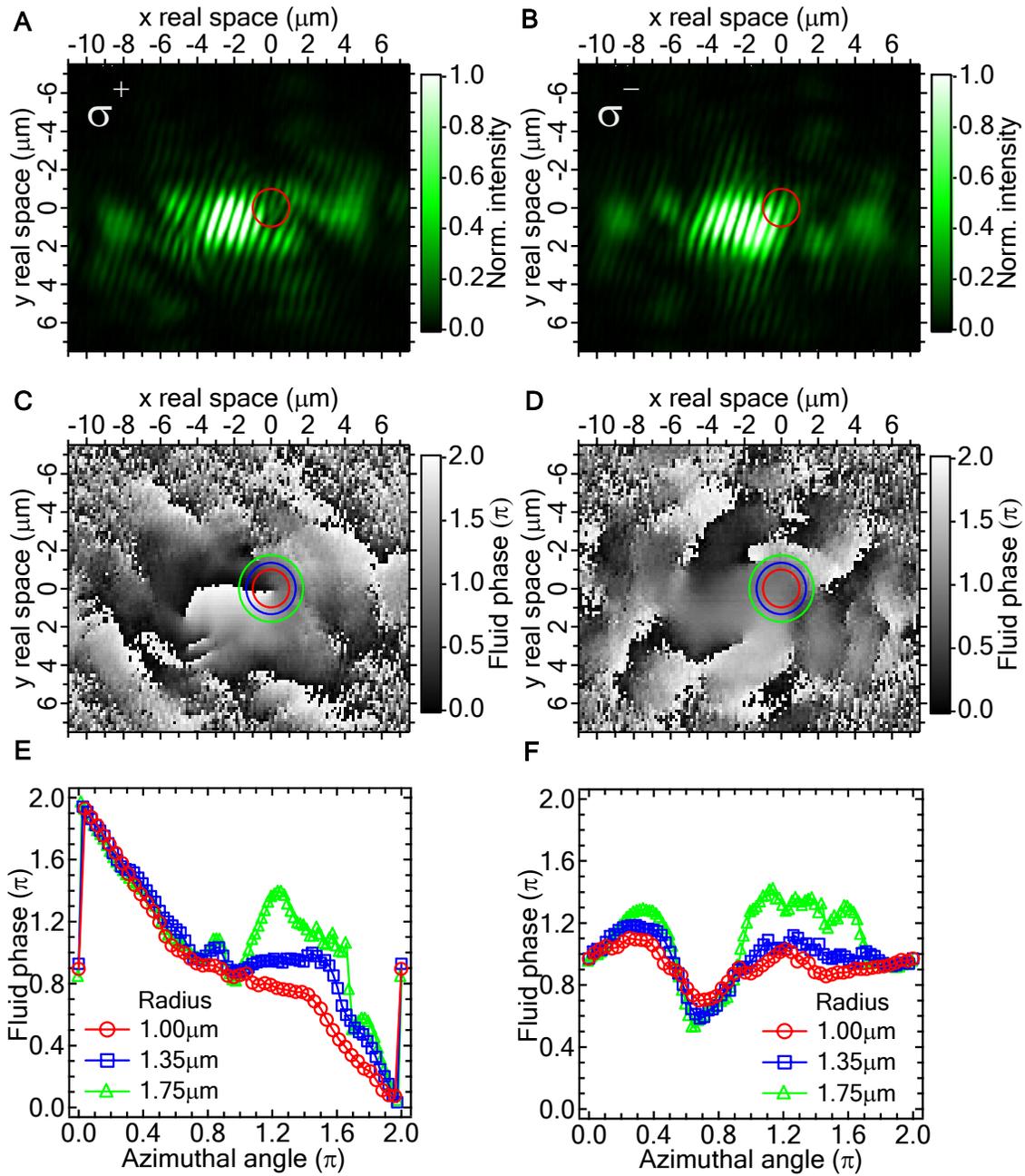

**Fig. 1**: (**A,B**): Reconstructed interferograms for $\sigma^+$ and $\sigma^-$ polarizations at the energy of the condensate. (**C,D**): Real space phase map extracted from the interferograms (**A**) and (**B**) for polarizations $\sigma^+$ and $\sigma^-$ respectively. The three circles with different colors denote the real space paths over which we have plotted the phase as a function of the azimuthal angle in panels (**E,F**). The phase profiles in $\sigma^+$ polarization (**E**) show that the phase changes by $-2\pi$ when circumventing the vortex core, which is the signature of the singularity. In contrast, for the $\sigma^-$ polarization (**F**), we see a quasi-flat phase profile with zero overall phase shift as one goes around the core. The farther we probe the phase away from the vortex core, the more the phase diverges from the linear behavior vs angle.

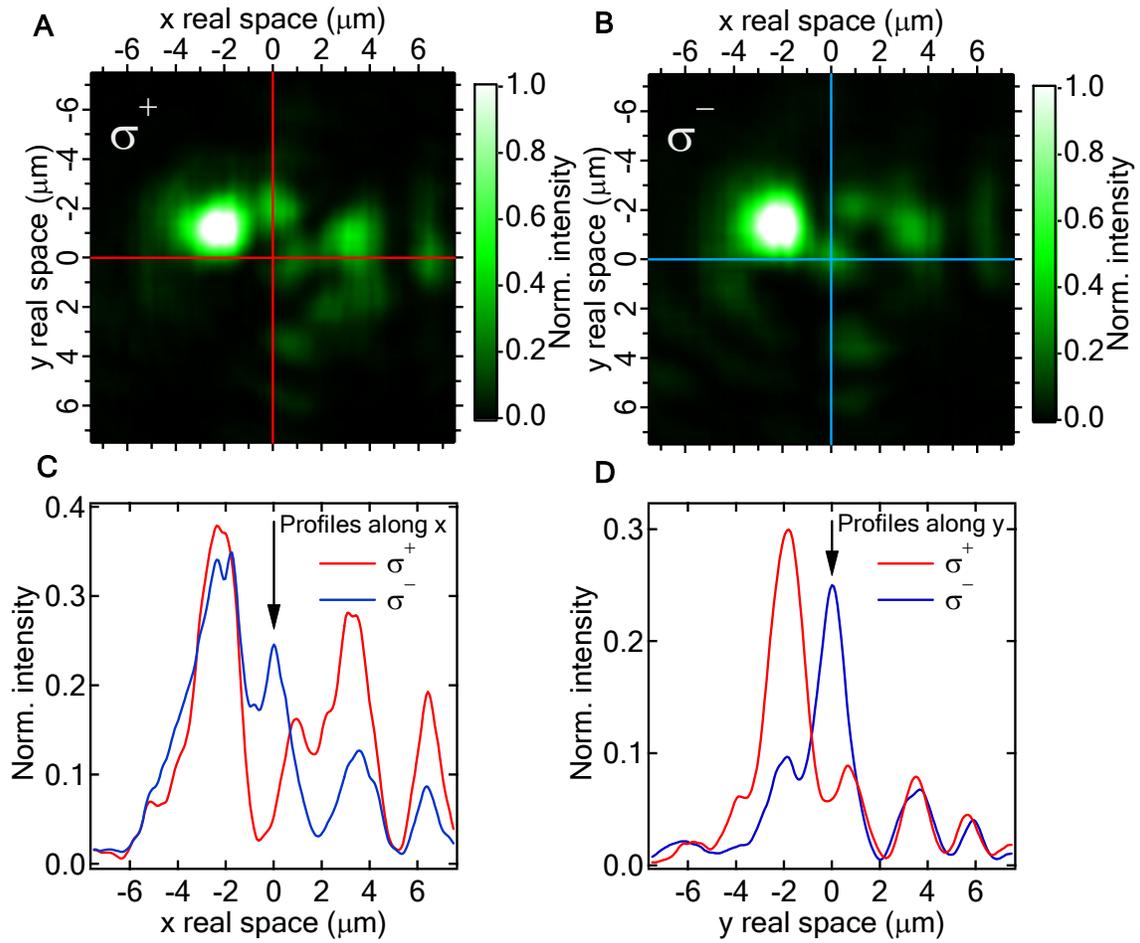

**Fig. 2**: (**A,B**): Polariton densities in real space at the frequency of the condensate for $\sigma^+$ and $\sigma^-$ polarizations, respectively. It is easily seen that at the position of the vortex core (0,0) which is indicated by the red cross for $\sigma^+$ polarization and the blue cross for the $\sigma^-$ polarization, there is a local minimum for $\sigma^+$ polariton density and a local maximum for the $\sigma^-$ density. (**C**): Density profiles along $x$ direction for the two polarizations and (**D**): the corresponding density profiles along $y$ direction. Experimentally measured polariton density behaves in excellent agreement with the theoretical prediction (Eq. (3) and Ref. (*21*)). This behavior is also evident from the fact that half vortices must be fully circularly polarized at the centers of their cores.

# Supplementary on-line material for

# Observation of Half Quantum Vortices in an Exciton-Polariton Condensate

K. G. Lagoudakis[1], T. Ostatnický[2], A.V. Kavokin[2,3], Y. G. Rubo[4], R. André[5], B. Deveaud-Plédran[1]

[1]IPEQ, École Polytechnique Fédérale de Lausanne (EPFL),1015 Lausanne, Switzerland.

[2]School of Physics and Astronomy, University of Southampton, Highfield, Southampton SO17 1BJ, United Kingdom.

[3]Marie-Curie Chair of Excellence, University of Rome II, 1, via della Ricerca Scientifica, Rome, 00133, Italy.

[4]Centro de Investigación en Energía, Universidad Nacional Autónoma de México, Temixco, Morelos, 62580, Mexico.

[5]Institut Néel, CNRS, Grenoble, France.

**Experimental setup**

The sample was excited in a quasi-CW non-resonant way with the wavelength of the laser being tuned to the first minimum of the reflectivity spectrum below the stopband to avoid heating. We used an electronically locked, single longitudinal mode Ti:Sapphire laser which excited the sample with an excitation spot of quasi top hat intensity profile and about 10μm diameter. The linewidth of the condensate emission we observed here was of the order of 30 μeV, close to the spectral resolution of our spectrometer and narrower than the average polarization splitting (*S1*). In addition, using a monomode laser, we observed an enhanced mode-synchronization effect (*S2*) which allowed formation of a single condensate in the regions of the sample characterized by low disorder. The collection of the luminescence was done by a high numerical aperture microscope objective (N.A=0.5) allowing collection of light within a cone of ±30º and providing a submicron real space resolution. All spectral studies were performed using a double 1m long monochromator with ~20μeV resolution. The setup is depicted in Fig. S1. We have added a λ/4 before the interferometer at 45° with respect to the polarization analyzer which in this case was a Wollaston prism with 20° polarization splitting angle, to allow for simultaneous polarization resolved interferometry in $\sigma^+$ and $\sigma^-$ polarizations. The polarization resolved interferograms were sent simultaneously on two different regions of the same CCD camera and thus the images of the two polarizations were recorded at once. The four different kinds of HQVs are shown in Figs. S2 and S3. On the upper panels one has the raw data and on the lower panels we provide only the fringes, having removed numerically the CW part of the interferograms. In Fig. S2 (A,B) one can see a close pair of $(k,m) = \left(-\frac{1}{2}, +\frac{1}{2}\right)$ and $(k,m) = \left(+\frac{1}{2}, +\frac{1}{2}\right)$ HQVs being in the red circle and blue box respectively. In Fig. S3 (A,B) there is another pair of $(k,m) = \left(-\frac{1}{2}, -\frac{1}{2}\right)$ and $(k,m) = \left(+\frac{1}{2}, -\frac{1}{2}\right)$ HQVs in the orange circle and pink box, respectively.

For the spectrally resolved studies, we have replaced the CCD of Fig. S1 with the double monochromator and the Wollaston prism has been replaced with a normal polarizer. The optical tomography is then performed by shifting the lens L and acquiring one spectrally resolved image for each lens displacement. For the figures shown in the paper, we acquired ~100 slices.

**Mutual coherence of the two circular polarization components**

In order to rule out the possibility of having two independent condensates, one in one circular polarization carrying a full vortex and another one in the other polarization with no vortex, we have performed an additional interferometric experiment to probe the mutual coherence between the two polarizations. In this way we probe that the two polarization components are coming from the same two component spinor condensate. For this purpose we have built a modified polarization-mixing Mach-Zehnder interferometer as depicted in Fig. S4. The two polarization components are sent through the two independent arms of the interferometer and in order to achieve interference at the output, one of the two components is rotated by means of a half wave plate in order to become co-polarized with the other component. The interference then reveals whether the two components are mutually coherent. In this experiment we provide two figures where a half vortex is imaged by two independent methods, the polarization resolved interferometry as in Fig. S5 (A), and the polarization mixing interferometry as in Fig. S5 (B). As seen in this figure, when making the two polarizations interfere, we clearly see an interference pattern which proves that the two polarizations are mutually coherent. The half vortex is easily distinguishable as a fork like dislocation at the top left corner and no forklike dislocation (straight interference fringes) at the symmetric position with respect to the autocorrelation point, the coordinates of which are extracted from Fig. S6.

**Simulation of the interference patterns**

All the observed interference patterns of HQVs are reproduced by the present theory for both experimental setups. We have considered a single vortex or a pair of vortices in a polariton fluid generated by a Gaussian laser beam with a half-width of 10μm. The winding numbers are denoted in the corresponding figure captions. We used Eq. (3) and the definitions of the radial functions $f$ and $g$ from (S3) in order to calculate the complex electric field amplitudes $E_+$ and $E_-$ of a single vortex, emerging in each of the respective circularly polarized components. The resulting intensity of light in the experimental configuration of Fig. S1 is then expressed as:

$$I_\pm(x,y) = \left| E_\pm(x,y) \exp\left[-i\left(K_x x + K_y y\right)\right] + E_\pm(x_0 - x, y_0 - y) \right|^2, \qquad (S1)$$

where we use the coordinate system relevant to the real image on the sample, $(x_0, y_0)$ is the effective position of the inversion centre of the retro-reflector in terms of the sample coordinates and $(K_x, K_y)$ is the wave vector which determines the inclination of the beams from the respective interferometer arms. In order to simulate the interference pattern of a sample with two vortices, we approximated the electric field in plane of the sample by a weighted superposition of the fields of two spatially separated single HQVs:

$$E_\pm = \frac{\ell_2}{\ell_1 + \ell_2} E_{1\pm} + \frac{\ell_1}{\ell_1 + \ell_2} E_{2\pm}, \tag{S2}$$

where $E_{j\pm}$ is the particular circular component of the electric field emerging from the vortex $j = 1,2$ and the symbols $\ell_{1,2}$ in the weighting functions denote distances from the cores of the vortices 1 or 2, respectively. Weighting of the electric field components in Eq. (S2) ensures continuity of the overall electric field and its derivatives.

The calculated interference patterns produced by the interferometer in both Figs. S2 and S3 for the two circular polarizations are shown in Fig. S7 (A,B). The simulated interference fringes in the geometry shown in Fig. S5 (B) are plotted in Fig. S8 (A). One can see that the experimental images of Fig. S5 (B) are reproduced and the vortex is clearly identified. The calculated plot in Fig. S8 (B) shows the profile of the intensities of two circularly polarized components of the polariton field at the HQV core, which appears to be in excellent agreement with the experimental data as well.

**Pumping power dependence**

The observed half vortices show a strong dependence on the excitation intensity. Below the condensation threshold they do not exist and they usually appear when the condensation threshold is crossed. Then they tend to disappear at high power above threshold. The HQV shown in the main part of the paper gets unpinned when the excitation intensity crosses a value of roughly 4.5 times the condensation threshold

($P\approx4.5\cdot P_{th}$). In Fig. S9 (A,B) we show the interferograms for excitation powers 2.5 times above the condensation threshold ($P=2.5\cdot P_{th}$) and 5 times above the threshold ($P=5.0\cdot P_{th}$). The forklike dislocation in $\sigma^+$ polarization is clearly distinguishable for $P=2.5\cdot P_{th}$ (Fig. S9 A), but it totally disappears at $P=5.0\cdot P_{th}$ (Fig. S9 B), demonstrating unpinning of the half vortex due to the screening of the static disorder potential.

**Vortex formation**

It has been shown (*S4*) that the vortices are formed spontaneously above the condensation threshold as a result of the mutual action of the non-uniform pumping and decay of polaritons. In a disorder free sample one would expect formation of vortex-antivortex pairs in each of the circular polarizations. These pairs are nothing but the HQV bound pairs [(-1/2,+1/2),(+1/2,-1/2)] and [(-1/2,-1/2),(+1/2,+1/2)] (*S3*). Simulations performed using the spin-dependent Gross-Pitaevskii equations showed that vortices in the opposite circular polarizations are separated and pinned to specific locations due to the combined effect of the disorder and spin-dependent polariton-polariton interactions *(S5)*. Propagation of the polariton fluxes at the early stage of formation of the condensate determines the winding numbers of the pinned vortices. Pinned vortices whose winding numbers do not vary in a large number of experimental realizations indicates that their formation dynamics is nearly identical in different experiments, and stochastic fluctuations of the order parameter are negligible for the determination of the steady state in the presence of vortices in our structure.

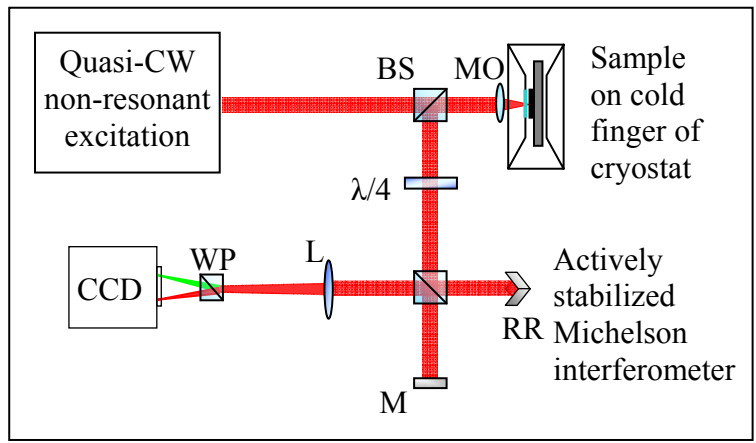

**Fig. S1**: Experimental setup of the polarization resolved interferometry. Two polarization components $\sigma^+$ and $\sigma^-$ get separated by means of the Wollaston prism and they are sent on the two opposite sides of the same CCD. In this manner we are facilitating the simultaneous imaging of both circular polarization components. BS: beam splitter, M: mirror, MO: microscope objective, $\lambda/4$: quarter wave plate, RR: retroreflector, L: lens, WP: Wollaston prism, CCD: charge-coupled device. In this setup both circular polarizations are propagating through both arms of the interferometer and only get separated by the time they traverse the Wollaston prism.

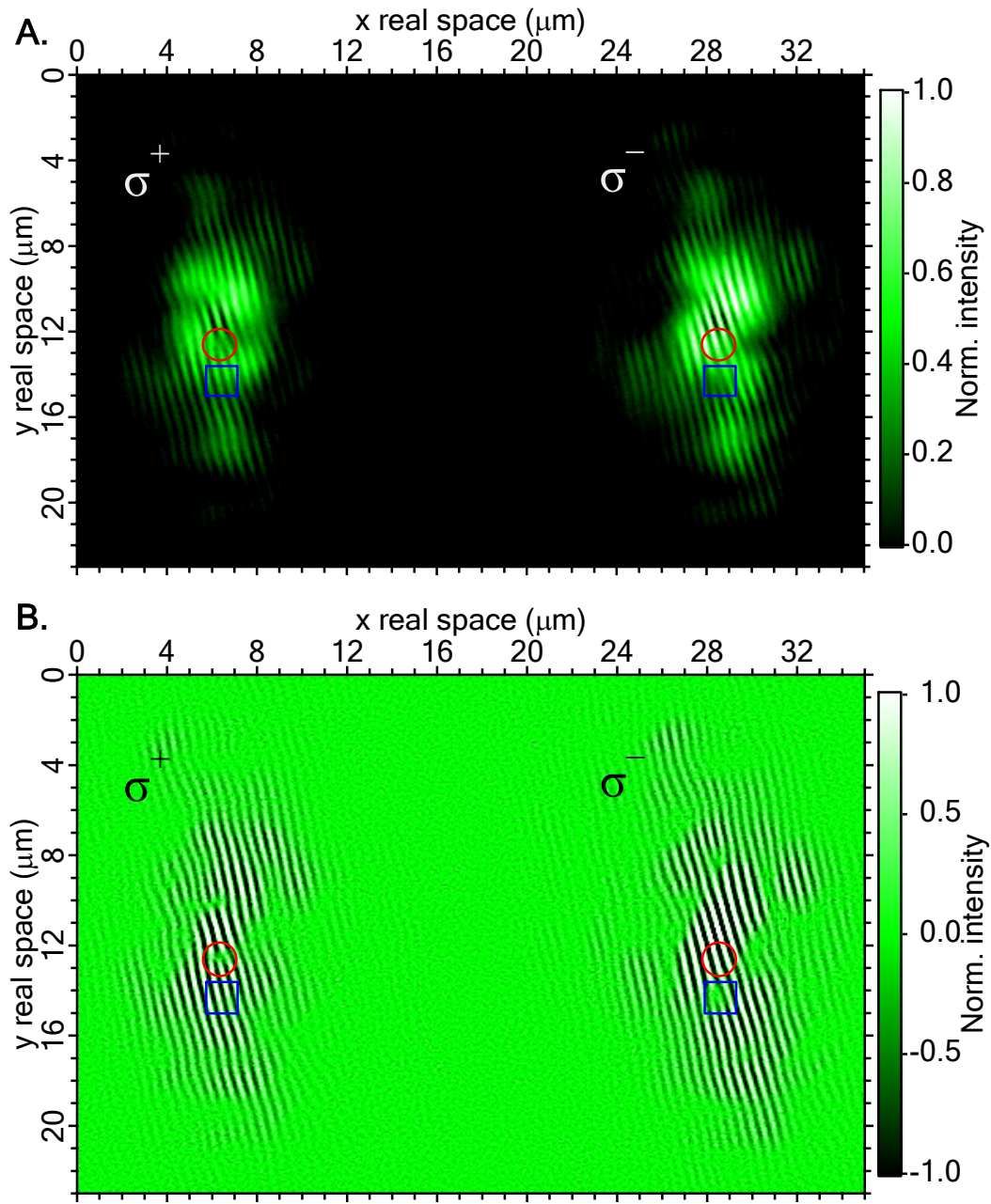

**Fig. S2**: Simultaneous imaging of both polarization components for the identification of half vortices. (**A**): Raw interferometric data and (**B**): interference fringes after removing numerically the CW part of the raw data interferogram. In this figure two independent HQVs are clearly distinguishable, one with winding numbers $(k,m) = \left(-\frac{1}{2}, +\frac{1}{2}\right)$ in the red circle and one with winding numbers $(k,m) = \left(+\frac{1}{2}, +\frac{1}{2}\right)$ in the blue box.

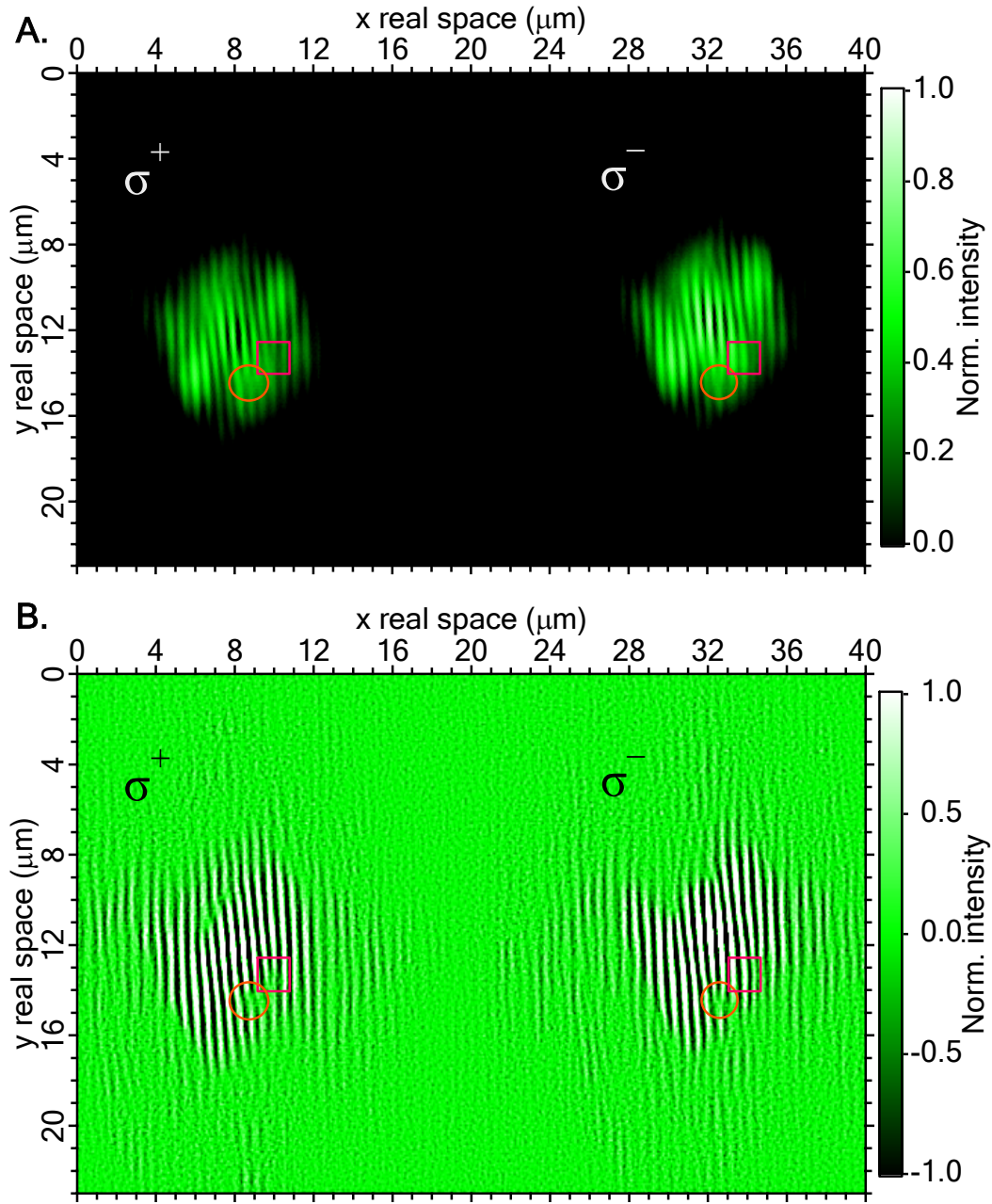

**Fig. S3**: Simultaneous imaging of both polarization components for the identification of half vortices at a different position on the sample showing another close pair of HQVs. (**A**): Raw interferometric data and (**B**): interference fringes after removing numerically the CW part of the raw data interferogram. In this figure two independent HQVs are also clearly distinguishable only here with winding numbers $(k,m)=\left(-\frac{1}{2},-\frac{1}{2}\right)$ and $(k,m)=\left(+\frac{1}{2},-\frac{1}{2}\right)$ in the orange circle and in the pink box respectively.

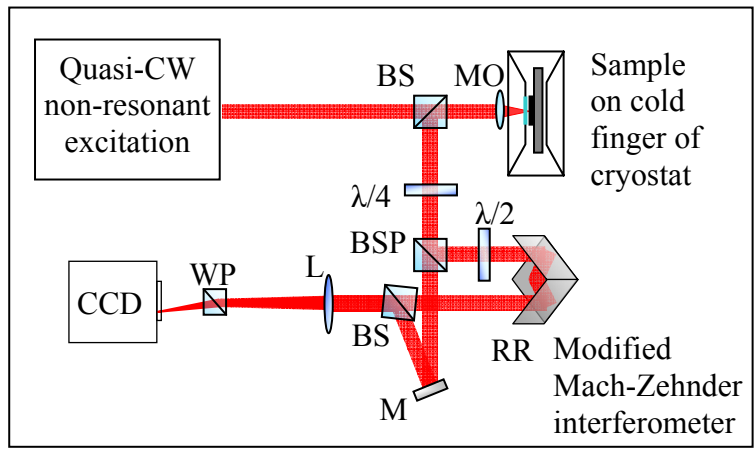

**Fig. S4**: Experimental setup of polarization mixing interferometry for the verification of the mutual coherence between the two circular polarization components. It consists of a polarization mixing Mach-Zehnder interferometer where the two polarizations get separated and sent on the two interferometer arms independently. The half wave plate is set at 45° and rotates one of the two polarizations by $\pi/2$ in order to coincide with the polarization in the other arm. The interference is then formed by mixing the two independent polarizations. BSP: beam splitter polarizer, $\lambda/2$: half wave plate.

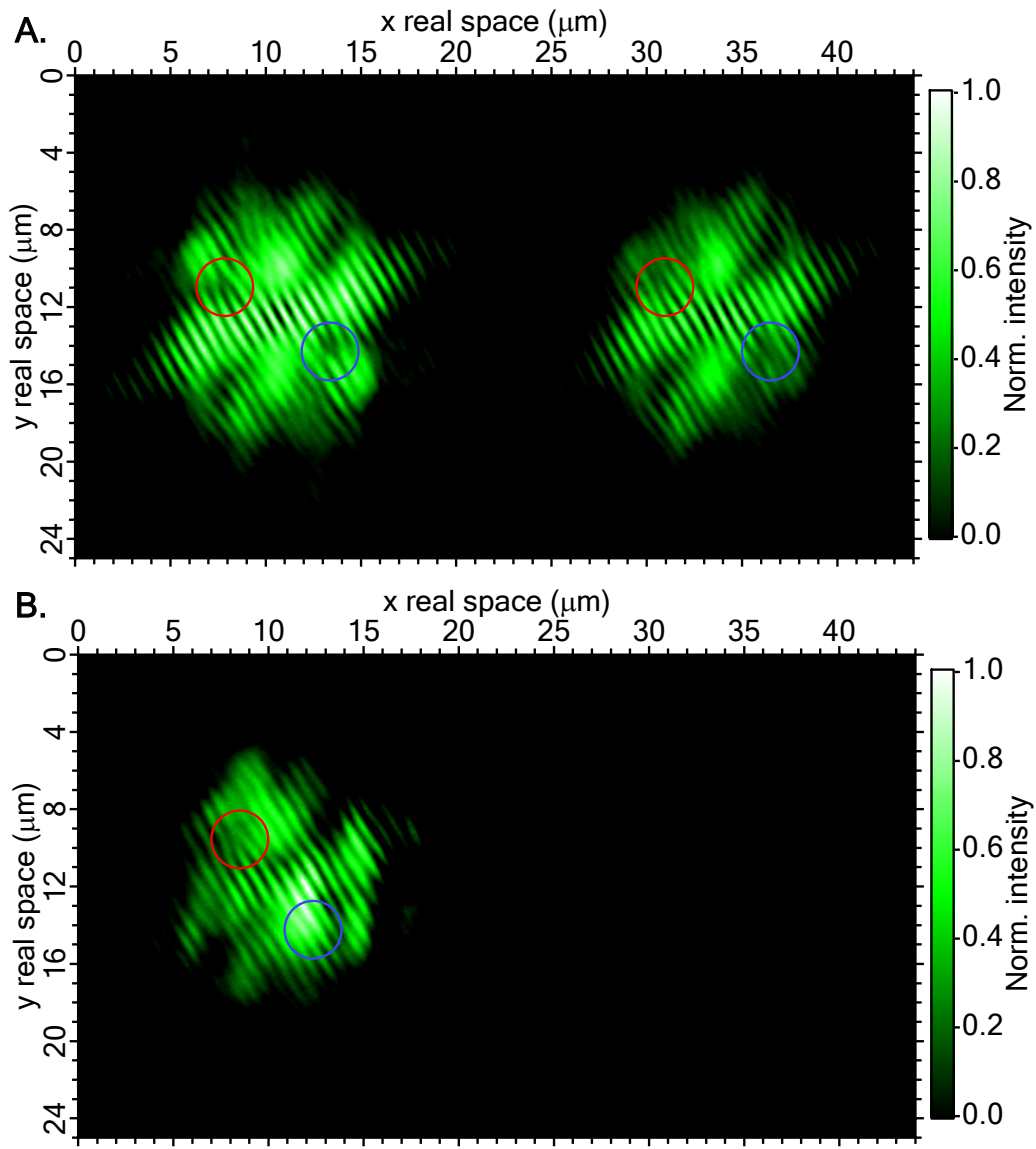

**Fig. S5**: (**A**): A half vortex observed with the standard method of simultaneous imaging of the interference patterns of the two polarization components. (**B**): The same HQV observed by means of polarization mixing interferometry. The half vortex shown in (**B**) is clearly observed as a forklike dislocation in the red circle and no forklike dislocation in the blue circle. The centers of the blue and red circles in both (**A**) and (**B**) are placed symmetrically with respect to the autocorrelation point. This experiment clearly demonstrates that the two polarization components are mutually coherent and that we can use whichever method facilitates our observations.

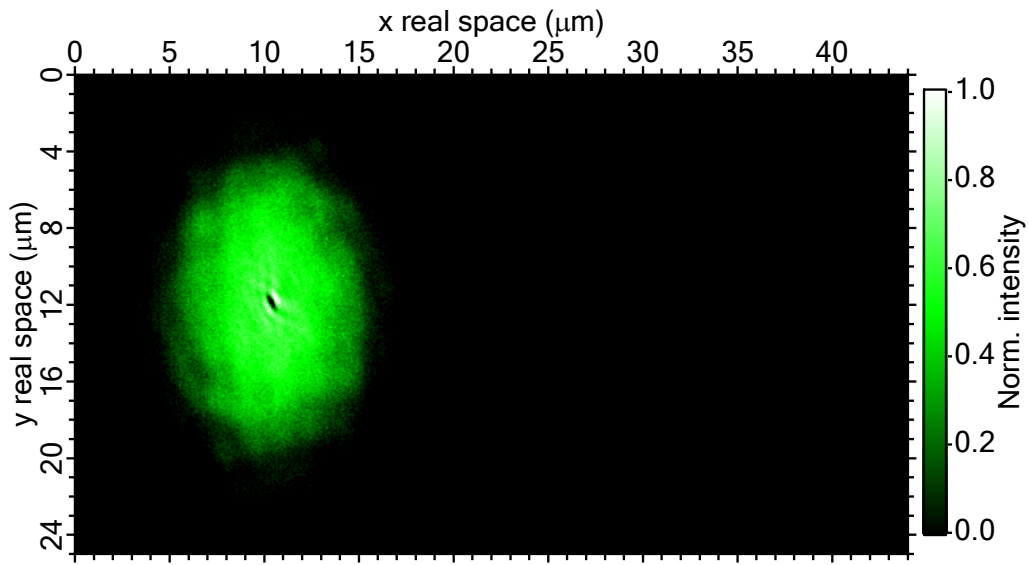

**Fig. S6**: Interference pattern below the condensation threshold as seen at the output of the Mach Zehnder polarization mixing interferometer. The autocorrelation point can be easily determined since it is the only place where the modulation of intensity caused by interference can be seen. Its size is defined by the response function of the microscope objective and the thermal de Broglie wavelength of polaritons. Here it is on a submicron scale. Note that the luminescence below threshold is not polarized thus we had to add a plate polarizer before the λ/4 to be able to see the interference at the autocorrelation point.

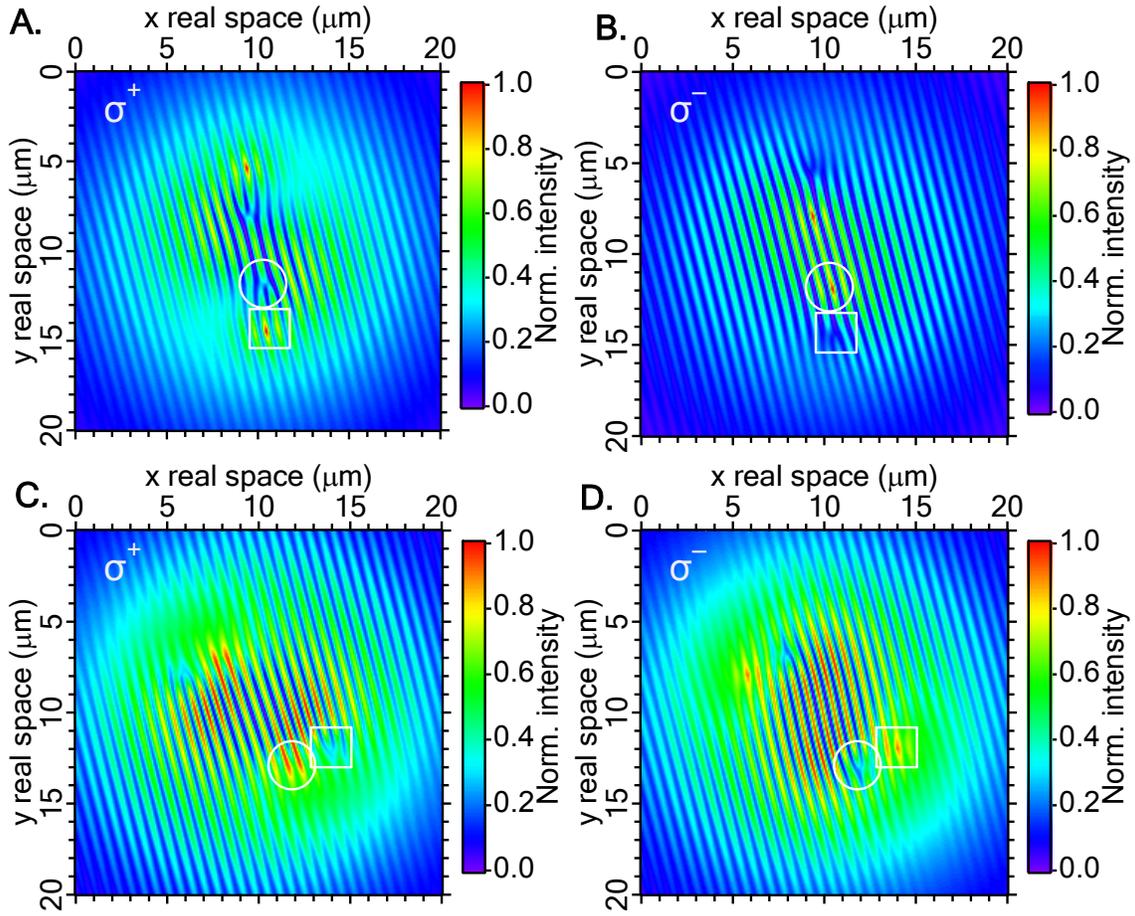

**Fig. S7**: Numerical simulation of the interference pattern in each of the circular polarizations for pairs of HQVs. (**A**): Simulated interferogram of a pair of HQVs with the same relative core coordinates as in figure S2. The quantum numbers are $(k,m) = \left(-\frac{1}{2}, +\frac{1}{2}\right)$ and $(k,m) = \left(+\frac{1}{2}, +\frac{1}{2}\right)$ in the circle and box respectively. The singularities behave identically to the experimental data. (**B**): Simulated interferogram of a pair of HQVs with the same relative core coordinates as in figure S3. The quantum numbers are here $(k,m) = \left(-\frac{1}{2}, -\frac{1}{2}\right)$ and $(k,m) = \left(+\frac{1}{2}, -\frac{1}{2}\right)$ in the circle and box respectively.

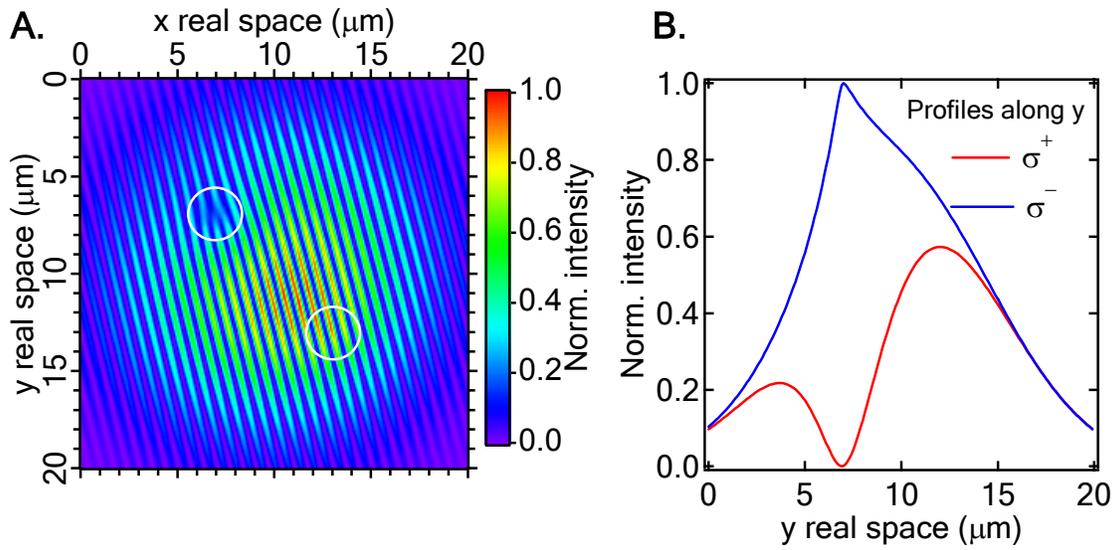

**Fig. S8**: (**A**) Simulated fringes emerging from interference of the two opposite circular components of the luminescence. The HQV with $(k,m) = \left(-\frac{1}{2}, +\frac{1}{2}\right)$ is situated at the coordinates (7,7) where a forklike dislocation is seen. (**B**) The calculated polariton field intensity across the vortex core in the direction of the *y* axis in two circular polarisations. At the center of the vortex the minimum in $\sigma^+$ polarisation coincides with the maximum in $\sigma^-$ polarisation.

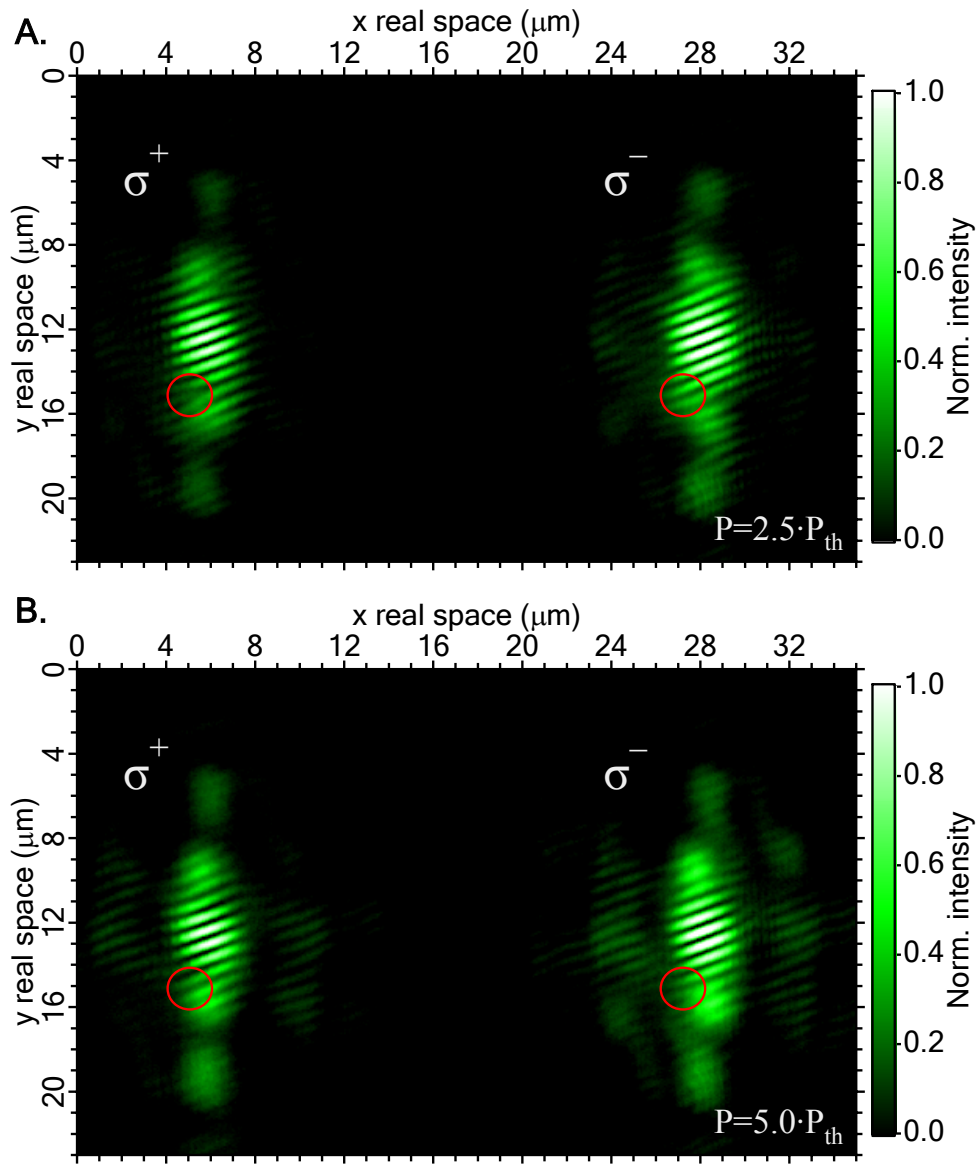

**Fig. S9**: Interference patterns for two excitation powers. In (**A**) the excitation was at a power 2.5 times higher that the condensation threshold whereas in (**B**) the excitation power was 5 times above the threshold. In panel (**A**) the half vortex is clearly shown by the red circles (forklike dislocation for sigma plus and straight fringes for sigma minus), whereas in panel (**B**) where the excitation power is high, in the same circles the interference pattern has changed with the most striking feature being the vanishing of the forklike dislocation. This clearly indicates the vortex unpinning from that specific location.